\documentclass[longauth]{aa}  

\usepackage{graphicx}
\usepackage{txfonts}
\usepackage{natbib}
\usepackage{verbatim}
\usepackage{amsmath}
\usepackage{xcolor}
\usepackage{txfonts}
\usepackage{keyval}
\usepackage{hyperref}

\def\beq{\begin{equation}}
\def\eeq{\end{equation}}
\def\iorb{i_{\rm obs}}
\def\oorb{\omega_{\rm obs}}
\def\Oorb{\Omega_{\rm obs}}
\def\sgra{SgrA$^*\,$}

\defcitealias{GRAVITY:2025ahf}{Paper I}
\newcommand{\LIRA}            {1}
\newcommand{\CotedAzur}      {2}
\newcommand{\MPE}             {3}
\newcommand{\IPAG}             {4}
\newcommand{\MPIA}             {5}
\newcommand{\UCologne}         {6}
\newcommand{\ESOGarching}    {7}
\newcommand{\USouthampton}    {8}
\newcommand{\KUL}    {9}
\newcommand{\FEUP}             {10}
\newcommand{\UBerkley}       {11}
\newcommand{\TUM}              {12}
\newcommand{\CENTRA}         {13}
\newcommand{\UCD}        {14}
\newcommand{\UNAM}        {15}
\newcommand{\CRAL}        {16}
\newcommand{\KAVLI}         {17}
\newcommand{\ESOSantiago}    {18}
\newcommand{\CNRSChile} {19}

\begin{document}

\title{Improving constraints on the Yukawa correction at the Galactic Center with multiple stellar orbits}

\author{
    The GRAVITY$^+$ Collaboration\fnmsep\thanks{
    GRAVITY is developed in collaboration by MPE / LIRA of Paris Observatory / CNRS / Sorbonne Université / Univ. Paris Diderot / IPAG of Université Grenoble Alpes / MPIA / Univ. of Cologne / CENTRA - Centro de Astrofisica e Gravitação / ESO. },
    A.~Foschi\inst{\LIRA}\fnmsep\thanks{Corresponding author: arianna.foschi@obspm.fr},
    K.~Abd~El~Dayem\inst{\LIRA},
    N.~Aimar\inst{\FEUP,\CENTRA},
    A.~Berdeu\inst{\ESOGarching,\LIRA},
    J.-P.~Berger\inst{\IPAG},
    G.~Bourdarot\inst{\MPE},
    W.~Brandner\inst{\MPIA},
    Y.~Cao\inst{\MPE},
    C.~Correia\inst{\FEUP,\CENTRA},
    S.~Cueves~Cardona\inst{\UNAM},
    R.~Davies\inst{\MPE},
    D.~Defr{\`e}re\inst{\KUL},
    F.~Delplancke-Str{\"o}bele \inst{\ESOGarching},
    A.~Drescher\inst{\IPAG},
    F.~Eisenhauer\inst{\MPE,\TUM},
    L.~Esteras~Otal\inst{\ESOGarching},
    M.~Fabricius\inst{\MPE},
    H.~Feuchtgruber\inst{\MPE},
    S.~Flesch\inst{\MPE},
    N.M.~F{\"o}rster~Schreiber\inst{\MPE},
    Q.~Fournier\inst{\MPE},
    P.~Garcia\inst{\FEUP,\CENTRA},
    R.~Garcia~Lopez\inst{\UCD},
    R.~Genzel\inst{\MPE,\UBerkley},
    S.~Gillessen\inst{\MPE},
    F.~Gont{\'e}\inst{\ESOGarching},
    X.~Haubois\inst{\ESOSantiago},
    S.F.~H{\"o}nig\inst{\USouthampton},
    M.~Houll{\'e}\inst{\IPAG},
    S.~Joharle\inst{\MPE},
    J.~Kammerer\inst{\ESOGarching},
    A.~Kaufer\inst{\ESOSantiago},
    P.~Kervella\inst{\LIRA, \CNRSChile},
    L.~Kreidberg\inst{\MPIA},
    L.~Labadie\inst{\UCologne},
    S.~Lacour\inst{\LIRA,\ESOGarching},
    O.~Lai\inst{\CotedAzur},
    R.~Laugier\inst{\KUL},
    J.-B.~Le~Bouquin\inst{\IPAG},
    J.~Leftley\inst{\USouthampton},
    R.~Li\inst{\MPE},
    B.~Lopez\inst{\CotedAzur},
    D.~Lutz\inst{\MPE},
    F.~Mang\inst{\MPE},
    A.~M{\'e}rand\inst{\ESOGarching},
    F.~Millour\inst{\CotedAzur},
    M.~Montarg{\`e}s\inst{\LIRA},
    N.~Moruj{\~a}o\inst{\FEUP,\CENTRA},
    H.~Nowacki\inst{\CotedAzur},
    M.~Nowak\inst{\LIRA},
    J.~Osorno\inst{\LIRA},
    T.~Ott\inst{\MPE},
    S.~Pappert\inst{\MPE},
    C.~Paladini\inst{\ESOSantiago},
    T.~Paumard\inst{\LIRA},
    K.~Perraut\inst{\IPAG},
    G.~Perrin\inst{\LIRA},
    R.~Petrov\inst{\CotedAzur},
    N.~Pourr{\'e}\inst{\IPAG},
    S.~Rabien\inst{\MPE},
    D.C.~Ribeiro\inst{\MPE},
    S.~Robbe-Dubois\inst{\CotedAzur},
    M.~Sadun~Bordoni\inst{\MPE},
    J.~Sanchez-Bermudez\inst{\UNAM},
    J.~Sauter\inst{\MPIA},
    J.~Scigliuto\inst{\CotedAzur},
    J.~Shangguan\inst{\MPE,\KAVLI},
    T.T.~Shimizu\inst{\MPE},
    F.~Soulez\inst{\CRAL},
    C.~Straubmeier\inst{\UCologne},
    E.~Sturm\inst{\MPE},
    M.~Subroweit\inst{\UCologne},
    C.~Sykes\inst{\USouthampton},
    L.J.~Tacconi\inst{\MPE},
    P.~Th{\'e}venet\inst{\LIRA},
    I.~Urso\inst{\LIRA},
    F.H.~Vincent\inst{\LIRA},
    J.~Woillez\inst{\ESOGarching},
    G.~Zins\inst{\ESOSantiago}
}

\institute{
    LIRA, Observatoire de Paris, Universit{\'e} PSL, CNRS, Sorbonne Universit{\'e}, Universit{\'e} de Paris, 5 place Jules Janssen, 92190 Meudon, France
    \and 
    Universit{\'e} C{\^o}te d'Azur, Observatoire de la C{\^o}te d'Azur, CNRS, Laboratoire Lagrange, France
    \and
    Max Planck Institute for Extraterrestrial Physics, Giessenbachstra{\ss}e 1, 85748 Garching, Germany
    \and
    Univ. Grenoble Alpes, CNRS, IPAG, 38000 Grenoble, France
    \and
    Max Planck Institute for Astronomy, K{\"o}nigstuhl 17, 69117 Heidelberg, Germany
    \and
    1st Institute of Physics, University of Cologne, Z{\"u}lpicher Stra{\ss}e 77, 50937 Cologne, Germany
    \and
    European Southern Observatory, Karl-Schwarzschild-Stra{\ss}e 2, 85748 Garching, Germany
    \and
    School of Physics \& Astronomy, University of Southampton, Southampton, SO17 1BJ, United Kingdom
    \and
    Institute of Astronomy, KU Leuven, Celestijnenlaan 200D, 3001 Leuven, Belgium
    \and
    Faculdade de Engenharia, Universidade do Porto, rua Dr. Roberto Frias, 4200-465 Porto, Portugal
    \and
    Departments of Physics \& Astronomy, Le Conte Hall, University of California, Berkeley, CA 94720, USA
    \and
    Technical University of Munich, TUM School of Natural Sciences, Physics Department, 85747 Garching, Germany
    \and
    CENTRA -- Centro de Astrof{\'i}sica e Gravita\c{c}{\~a}o, IST, Universidade de Lisboa, 1049-001 Lisboa, Portugal
    \and
    School of Physics, University College Dublin, Belfield, Dublin 4, Ireland
    \and
    Universidad Nacional Aut\'onoma de M\'exico. Instituto de Astronom\'ia. A.P. 70-264, 04510, Ciudad de M\'exico, M\'exico
    \and
    Univ. Lyon, Univ. Lyon 1, ENS de Lyon, CNRS, Centre de Recherche Astrophysique de Lyon UMR5574, F-69230 Saint-Genis-Laval, France
    \and
    Kavli Institute for Astronomy and Astrophysics, Peking University, Beijing, China
    \and
    European Southern Observatory, Casilla 19001, Santiago 19, Chile
    \and
    French-Chilean Laboratory for Astronomy, IRL 3386, CNRS and U. de Chile, Casilla 36-D, Santiago, Chile.
}

   \date{
   }

 
  \abstract
   {}
   {We investigate the presence of a Yukawa-like correction ($\propto \, \alpha e^{- r/\lambda}$)  to Newtonian gravity at the Galactic Center, using a multi-star fitting code, including the newly discovered star S$301$, to improve the constraints obtained using S$2$ orbit alone.}
   {We perform a Markov Chain Monte Carlo analysis using the astrometric and spectroscopic data of stars S$2$, S$55$, S$29$, S$38$ and S$301$ collected by GRAVITY, GRAVITY$^+$, NACO and SINFONI instruments, covering the period from $1992$ to $2025$.}
   {Compared to GRAVITY Collaboration 2025 (Paper I), in which only S$2$ was fitted, the tightest bound on the Yukawa coupling is again reached at $\lambda = 3\cdot10^{13}\,\rm m\ (\sim 200\,AU)$, where we find $|\alpha| < 6 \cdot10^{-4}$, an improvement of roughly a factor five. The addition of S$301$ extends the constraint to short ranges that were inaccessible to S$2$, yielding $|\alpha| < 0.004$ at $\lambda =10^{12}\,\rm m$ and $|\alpha| < 0.006$ at $\lambda =5\cdot 10^{11}\,\rm m$, while S$29$, with its larger apoapsis, gives $|\alpha| < 0.1$
at $\lambda \sim 2\cdot10^{15}\,\rm m$. The latter two represent the tightest limits at these scale lengths on a Yukawa correction obtained to date around a supermassive black hole.}
   {}

   \titlerunning{Yukawa correction at the Galactic Center with multiple stellar orbits}
   \authorrunning{GRAVITY$^+$ Collaboration: A. Foschi et al.}
   \maketitle
     
%

\section{Introduction}
General Relativity (GR) remains the successful theory of gravity, validated at Solar System scales and by gravitational waves and binary pulsars \citep{Will:2018bme, Nitz:2021uxj}, yet its shortcomings, like the unexplained cosmological constant \citep{Weinberg:1988cp, Peebles:2002gy}, the dark matter problem \citep{ Massey:2010hh, Salucci:2018hqu}, and the absence of a quantum completion \citep{Esposito:2011rx, Kiefer:2023bld}, still motivate the search for deviations. 

Among the possible modifications, a Yukawa-like correction to the Newtonian potential is particularly fascinating because it arises generically in the weak-field limit of a broad class of Extended Theories of Gravity (ETGs) \citep{Hoyle:2000cv, Moffat:2005si, Hinterbichler:2011tt, Alsing:2011er, Capozziello:2015lza}, as well as in dark matter (DM) models with a light mediator \citep{Gradwohl:1992ue, Carroll:2008ub}. We refer to \citetalias{GRAVITY:2025ahf} and \cite{deLaurentis:2022oqa} for a broader discussion on these topics.

The fifth force strength $\alpha$ is tightly constrained in the Solar System \citep{Hofmann:2018, MICROSCOPE:2022doy, Fienga:2023ocw, Tsai:2023zza}, but many of the theories predicting it also invoke a screening mechanism that suppress the effect precisely in low-curvature regimes like the Solar System. Thus, the GC probes a different environment: constraints there have been derived from the \sgra shadow \citep{Vagnozzi:2022moj}, from the Schwarzschild precession of S2 \citep{GRAVITY:2020gka, Jovanovic:2022twh}, and from analysis of publicly available or mock S-star data \citep{Zakharov:2018cbj, deMartino:2021daj, DellaMonica:2021xcf}. 

In \citetalias{GRAVITY:2025ahf} we performed a full fit of the astrometric and spectroscopic data of S2 collected by GRAVITY, NACO and SINFONI from $1992$ to $2023$, obtaining the strongest bound to date near a supermassive black hole (SMBH) at scales comparable to the S2 - \sgra separation ($\lambda \sim 3 \cdot 10^{13}$ m).

The Yukawa correction is controlled by two parameters, the intensity $\alpha$ and the scale length $\lambda$, and a star with pericenter $r_p$ and apocenter $r_a$ is sensitive only to $\lambda$ comparable to the radii it samples: for $\lambda \ll r_p$ the fifth force is exponentially suppressed along the whole orbit, while for $\lambda \gg r_a$ it becomes degenerate with a rescaling of the central SMBH mass.

In this paper we combine astrometric and, when available, spectroscopic data of S$2$ with those of S$55$, S$38$, S$29$ and the newly discovered S$301$ \citep{GRAVITY:2026}. Because the S$2$ data still dominate the fit, the first three stars mainly reduce the statistical uncertainty on $|\alpha|$, tightening the constraint over the same range of scale lengths probed by S$2$ alone ($\lambda \sim 10^{13}-10^{14}\rm \, m$). 

However, S$301$, whose pericenter passage is roughly $10$ times smaller than that of S$2$, allow us to constrain $|\alpha|$ in a short-range regime ($\lambda \sim 10^{11}-10^{12}\, \rm m$) that was previously inaccessible. In contrast, the larger apoapsis of S$29$ improves the upper limits at $\lambda \gtrsim 10^{15}\, \rm m$.


\section{Observations}
In this paper we include the data collected at the VLT by NACO, SINFONI, GRAVITY, the Enhanced Resolution Imager and Spectrograph (ERIS)~\citep{Davies:2023}, operational at VLT from 2022, after the decommissioning of SINFONI, and GRAVITY$^+$. The latter is an on-going GRAVITY upgrade, with its new adaptive optics and laser guide stars that enable fainter GC observations \citep{Gravity:2022, 2026A&A...707A.115G}. 
\begin{itemize}
        \item For S$2$ we used 129 NACO and 82 GRAVITY points for the astrometry; 92 SINFONI and 3 ERIS data points for the radial velocity measurements. These data cover the time span $1992-2022$.
        \item For S$29$ we used 66 NACO and 29 GRAVITY data points for the astrometry and 17 SINFONI and 2 GNIRS radial velocity measurements. These data cover span from $2002$ to $2022$.
        \item For S$38$ we used 110 NACO and 23 GRAVITY data points for the astrometry and 8 SINFONI and 1 ERIS radial velocity measurements. These data spans from $2004$ to $2022$.
        \item For S$55$ we used 42 NACO and 27 GRAVITY points for the astrometry and 2 SINFONI measurements for the radial velocity, covering the time span $2004-2022$.
        \item For S$301$ we used 19 astrometric points collected by GRAVITY$^+$ between $2023.3$ and $2025.4$. 
       To complete the orbit, we produced a mock data set with 13 points between $2017.4$ and $2023.3$, to cover the periastron passage, and 9 points between $2025.9$ and $2030$, for a total of 41 astrometric points. 
        These mock data are generated following the orbit described in \cite{GRAVITY:2026} with uncertainties that average around $\sigma \approx 50 \rm\,  mas$. 
        Radial velocity measurements are still not available for S$301$ and thus they are not included in the fit. As a sanity check, we test that the inclusion of a mock dataset for radial velocity does not alter the results. 
\end{itemize}

\section{Methods}
\label{Method}

The potential investigated has the following form:
\beq
U = -\frac{G M}{r} \left(1 + |\alpha| e^{-r/\lambda}\right)\, , 
\label{yukawa}
\eeq
where $\alpha$ represents the strength of interaction and $\lambda$ is a scale parameter which depends on the specific theory considered. 

If, for instance, we include a new massive field in the theory, $\lambda$ represents its Compton wavelength, which is related to its physical mass by $m_{\varphi} = h/c \lambda$, where $h$ is the Planck constant. 

The numerical integration of the equations of motion is performed using a Python code that implements the methods of Osculating Orbits in General Relativistic Environments (OOGRE) developed by the LIRA group at Paris Observatory \citep{Heissel:2021pcw}.

The latter is based on the perturbed Kepler model for which the variation of the osculating elements $\mu_i = \{p, e, \oorb, \Oorb, \iorb\}$, describing the star's orbit, is given by 
\begin{align}
& \frac{d p}{dt}  = 2 \sqrt{\frac{p^3}{G M_{\bullet}}}\frac{1}{(1-e \cos f)^3} \mathcal{S},
\label{dpdf} \\
&\frac{de}{dt}  = \sqrt{\frac{p}{G M_{\bullet}}} \left[ \sin f \, \mathcal{R} +  \frac{2 \cos f + e(1 + \cos^2 f)}{1 + e \cos f} \mathcal{S} \right] ,
\label{dedf}
\end{align}

\beq
\begin{split}
\frac{d \oorb}{dt}  = & \frac{1}{e} \sqrt{\frac{p}{G M_{\bullet}}} \left[- \cos f\, \mathcal{R} + \frac{2 + e \cos f}{1 + e \cos f} \, \sin f \, \mathcal{S} \right. \\
& \left. - e \cot \iorb \frac{\sin(\oorb + f)}{1 + e \cos f} \mathcal{W} \right]\, ,
\label{domegadf}
\end{split}
\eeq
\begin{align}
&\frac{d \iorb}{dt}  = \sqrt{\frac{p}{G M_{\bullet}}} \frac{\cos(\oorb + f)}{1 + e \cos f} \mathcal{W} \, , \\
&\frac{d \Oorb}{dt} = \sqrt{\frac{p}{G M_{\bullet}}} \frac{\sin(\oorb + f)}{1 + e \cos f} \frac{\mathcal{W}}{\sin \iorb} \, ,
\label{dOmegadf}
\end{align}
where the three components $\mathcal{R}$, $\mathcal{S}$ and $\mathcal{W}$ represent the acceleration defined in the frame co-moving with the star and $f$ is the true anomaly. The latter is identified by the Gaussian frame $(\mathbf{n}, \mathbf{l}, \mathbf{z})$, where $\mathbf{n} = \mathbf{r}/r$ is the radial vector, $\mathbf{z}$ is aligned with the star's angular momentum axis, and $\mathbf{l}$ completes the triad. 

Since the potential in Eq.~\eqref{yukawa} is spherically symmetric, the only component to be included is the radial one, i.e.,
\beq
\mathcal{R}_{\rm Y} = -\frac{G M_{\bullet}}{p^2 \lambda} e^{-p/(\lambda (1 + e \cos f))} \alpha (1 + e \cos f) (p + \lambda (1 + e \cos f )),
\eeq
where we have used the fact that $r = p/(1 + e \cos f)$. 

Together with the Yukawa-correction we also include in our fit the Schwarzschild precession \citep{GRAVITY:2020gka, GRAVITY:2024tth}, using the first order Post Newtonian (PN) acceleration, which in the co-moving frame of the star can be decomposed in the two components
\begin{align}
    & \mathcal{R}_{\rm 1PN} =  \frac{G^2 M_{\rm \bullet}^2}{c^2 p^3} ( 1+e \cos f)^2 \left(3(e^2+1)+2 e \cos f  - 4 e^2 \cos^2 f
    \right),\\
     & \mathcal{S}_{\rm 1PN} = \frac{4 G^2 M_{\bullet}^2}{c^2 p^3} \left(1 + e \cos f\right)^3 e \sin f.
\end{align}
As already stated in \citetalias{GRAVITY:2025ahf}, a formal parametrized Post Newtonian (PN) treatment is not possible when a massive field is included in the action \citep{Alsing:2011er, PoissonWill2012}. 
This means that the model considered in this analysis is valid only for those ETGs that are indistinguishable from GR at $1$PN order, while theories that have parametrized PN parameters significantly different from unity are not considered \citep{Will:2014kxa}.

When integrating the equations of motion, other effects are included: the R\o mer delay, the relativistic Doppler shift and the gravitational redshift, the motion of the Solar System. All these effects are summarised in Appendix \ref{app:rel_effects}

Once the variation of the orbital elements is computed, the resulting osculating orbit must be compared against the data, which are collected into the observer frame, $\{\mathbf{X}, \mathbf{Y},\mathbf{Z}\}$. The transformation between the orbital plane, identified by $\{\mathbf{x}_{\rm orb}, \mathbf{y}_{\rm orb}, \mathbf{z}_{\rm orb}\}$ and the observer frame is reported in Appendix \ref{app:coord_transf}. 

To determine the best fit values, we perform an MCMC analysis with the Python package \textsc{emcee}~\citep{ForemanMackey:2013}, adopting $N_{\rm steps} = 5000$, $N_{\rm walkers} = 100$ and an initial burn-in phase of $ N \sim 500$ steps, that we set after inspecting the trace plots, depending on the model under consideration. We use a single Gaussian log-likelihood that joins the datasets of all stars, and uniform priors; the initial guesses of the parameters are reported in Appendix~\ref{app:initial_values}.
The convergence of the MCMC analysis is assessed by means of the autocorrelation time $\tau_c$, that is, we ensure that the chain length satisfies $N_{\rm steps} \gtrsim 50 \, \tau_c$.

During the fitting procedure, the Yukawa length scale $\lambda$ is held fixed across the range $10^{11} \leq \lambda \leq 10^{16} \, \rm m$, while we solve for the best-fit intensity $\alpha$ together with the remaining parameters characterizing each star.

Specifically, these parameters comprise the osculating elements $\mu_i$ of each star, along with the shared parameters $M_{\bullet}$, the mass of \sgra, $R_0$, the Galactic Center distance, and the offsets $\{x_0, y_0, v_{x0}, v_{y0}, v_{z0}\}$, which align the various datasets onto a common origin centered on \sgra~\citep{2015MNRAS.453.3234P}. 

The upper limit on $\alpha$ is extracted from its posterior distribution, which in most cases is described by a Gaussian for which we quote the $2 \sigma$ bound, corresponding to $\approx 95\%$. In the regime where $\alpha$ becomes difficult to sample, its posterior tends to flatten and the upper limit is obtained from the value enclosing $95\%$ of the area under the posterior.

\section{Results}
\label{sec:results}

\begin{figure}
   \centering  \includegraphics[width=\hsize]{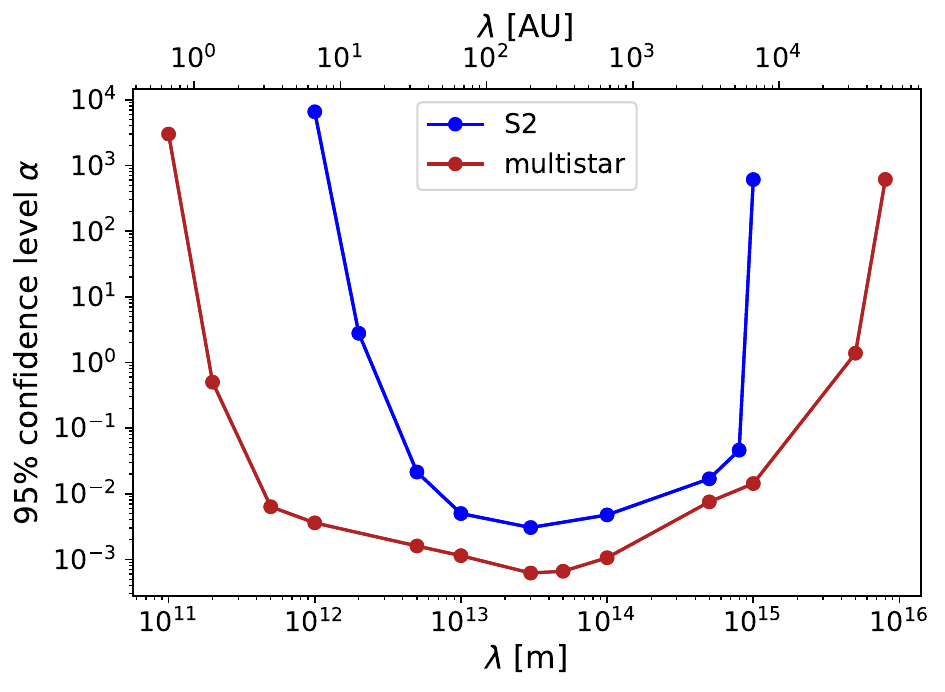}
      \caption{$95\%$ confidence level on $|\alpha|$ as function of the scale length $\lambda$. The red curve represents the upper limits when S$2$, S$55$, S$29$, S$38$ and S$301$ are included in the fit, while the blue curve is obtained using S$2$ only (from \citetalias{GRAVITY:2025ahf}).}
         \label{fig:conf_int}
\end{figure}

In Figure \ref{fig:conf_int} we show the $95\%$ confidence level on $|\alpha|$, as a function of $\lambda$ and we compare it with previous estimates obtained using only S$2$ \citepalias{GRAVITY:2025ahf}. 

Compared to the case with S$2$ alone, we observe that the presence of multiple stars improves the current best upper limit, found again at $\lambda = 3 \cdot 10^{13} \rm \, m$, by a factor 5, finding $|\alpha| < 0.0006$ and giving a new best bound for $|\alpha|$ at the GC. This is a consequence of S$2$ data dominating the $\chi^2$, with other stars that simply reduce the statistical uncertainty in the range $\lambda \sim 10^{13}-10^{14}\, \rm m$. 

The addition of S$29$, which has an apoapsis roughly $8$ times larger than S$2$, allows us to reduce the uncertainty on $|\alpha|$ around $\lambda \gtrsim 10^{15}\,\rm m$, a region that could not be sampled with S$2$ alone, due to the high degeneracy between $M_{\bullet}$ and $\alpha$. Here we obtain $|\alpha| < 0.1$ for $\lambda \sim 2 \cdot 10^{15}\,\rm m$, while for larger $\lambda$ the Yukawa term is absorbed into a rescaling of $M_{\bullet}$ and the constraints progressively degrade.  

However, the main novelty of this analysis is represented by the inclusion of S$301$, which, due to a pericenter passage roughly ten times smaller than that of S$2$, samples radii small enough that the Yukawa term remains unsuppressed even for shorter $\lambda$, inaccessible to all previously analysed stars. 
We find $|\alpha| < 0.004$ at $\lambda = 10^{12}\,\rm m$ and $|\alpha| < 0.006$ at $\lambda = 5 \cdot 10^{11}\,\rm m$, improving the previous estimate at these scale lengths by several orders of magnitude. For $\lambda \lesssim 10^{11}\,\rm m$ the exponential term is suppressed and no meaningful constraint can be obtained.

Although Solar System tests remain much more stringent at comparable scale lengths (e.g. $|\alpha| \lesssim 10^{-10}$ at $\lambda \sim 10^{11}\, \rm m$), this is the tightest bound obtained to date around a SMBH, an environment in which the screening mechanisms invoked by many of these theories may behave differently, and which Solar System experiments cannot access by definition.

\begin{figure*}
   \centering  \includegraphics[width=\textwidth]{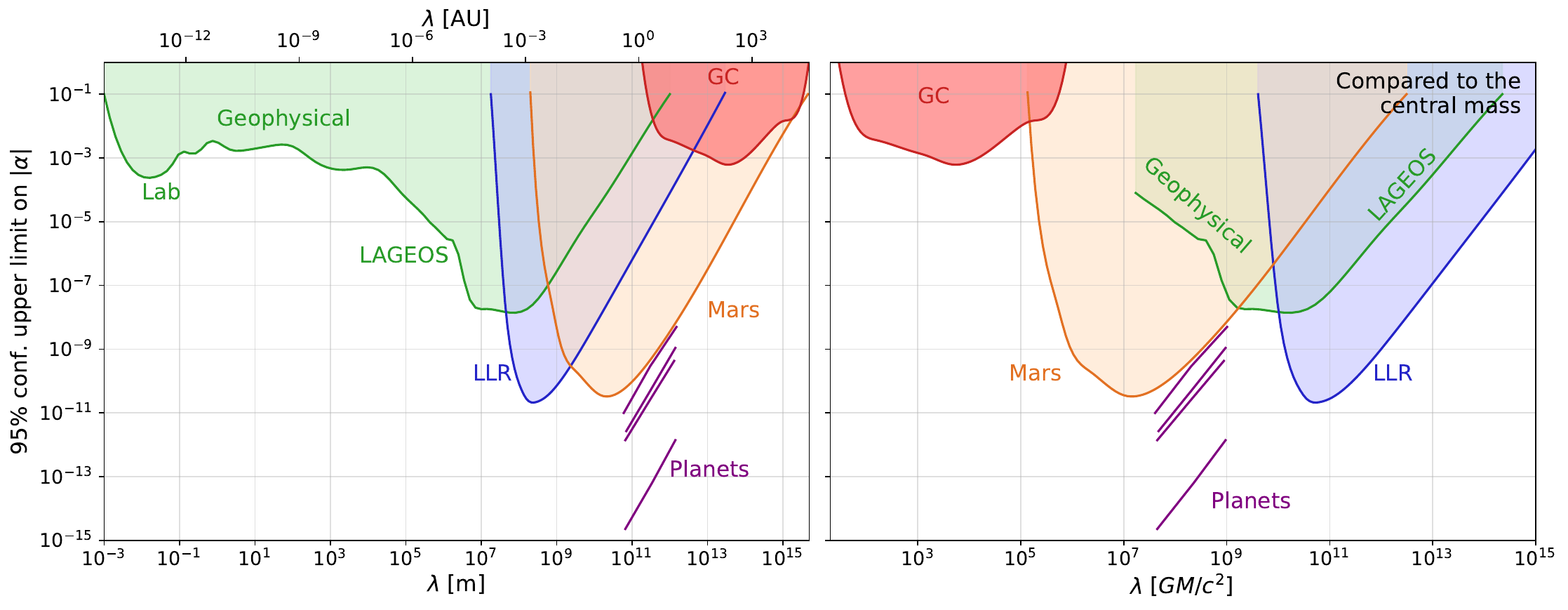}
      \caption{$95\%$ confidence level on $|\alpha|$ as function of the scale length $\lambda$, update of Figure 2 of \cite{Hees:2017aal}. The coloured regions are excluded by different experiments (see \cite{Konopliv:2011}, \cite{Fienga:2023ocw} and references therein), including the results reported in this work (red region). In the right panel the same constraints are reported in terms of the distance from the central mass generating the potential, showing the importance of the GC.}
         \label{fig:conf_int_region}
\end{figure*}

In Figure \ref{fig:conf_int_region} we plot the regions of $\alpha$ excluded by different experiments (from \cite{Konopliv:2011}), including the results at the GC obtained in this work, updating Figure 2 of \cite{Hees:2017aal}. In the right panel the length scale $\lambda$ is expressed in terms of the gravitational radius $G M/c^2$ of the central mass that generates the potential. The results obtained at the GC show the importance of testing this correction around \sgra, due to the closeness of the S-stars to the central BH.

In Appendix \ref{app:mock_vs_real} we discuss the need for mock data, and specifically points before the periastron passage, to complete the sampling of S301's orbit and thus obtain meaningful constraints at small values of $\lambda$.

\section{Conclusions}
In this paper, we update the current constraints on the fifth force intensity derived in \citetalias{GRAVITY:2025ahf} using a combined dataset that includes astrometric and spectroscopic data of S$2$, S$55$, S$29$, S$38$ and the newly discovered S$301$. 

We find an overall improvement over the entire range of scale lengths tested. The tightest bound on $|\alpha|$ is obtained at $\lambda = 3 \cdot 10^{13}\, \rm m$, as it was for S$2$, since the latter dominates the $\chi^2$ in that region. 

Thanks to the inclusion of multiple stars we improve the bound to be $|\alpha|< 0.0006$, a factor 5 more stringent than the previous one (cf. \citetalias{GRAVITY:2025ahf}). 

As mentioned in \citetalias{GRAVITY:2025ahf}, the inclusion of stars with different orbital range than S$2$ can improve the upper limit on $|\alpha|$ over a larger range of scale lengths. 

Indeed, thanks to S$29$, which has apoapsis larger than S$2$, we are able to constrain $|\alpha| \lesssim 0.1$ at $\lambda \sim 2 \cdot 10^{15} \rm\,  m $, i.e. where $\alpha$ was previously degenerate with a rescaling of the central mass $M_{\bullet}$. 
On the contrary, when including S$301$, which has a much smaller periapsis than S$2$, we significantly improve, by several orders of magnitude, the previous upper limits finding $|\alpha| < 0.004$ at $\lambda = 10^{12}\, \rm m$ and $|\alpha| < 0.006$ at $\lambda = 5 \cdot 10^{11}\, \rm m$.

\begin{acknowledgements}
 We are very grateful to our funding agencies (MPG, ERC, CNRS [PNCG, PNGRAM], DFG, BMBF, Paris Observatory [CS, PhyFOG], Observatoire des Sciences de l'Univers de Grenoble, and the Funda\c c\~ao para a Ci\^encia e a Tecnologia), to ESO and the Paranal staff, and to the many scientific and technical staff members in our institutions, who helped to make NACO, SINFONI, and GRAVITY a reality. This work was supported by Paris Île-de-France Region, the French National Research Agency (ANR) under grant ANR-23-EDIR-0003 (GRAFITY), the "Action Thématique Gravitation Références Astronomie Métrologie" (ATGRAM), the "Action Thématique Phénomènes Extrêmes et Multimessagers" (ATPEM), and the "Action Thématique Cosmologie et Galaxies" (ATCG), of
CNRS/INSU, with co-funding by CNRS/IN2P3, CNRS/INP, CEA and CNES. J.S. acknowledge the National Science Foundation of China (12233001) and the National Key R\&D Program of China (2022YFF0503401). J.S-B. acknowledges the support received by the UNAM DGAPA-PAPIIT project AG-101025 and from the SECIHTI Ciencia de Frontera project CBF-2025-I-3033. C.C., N.A., N.M. and P.G. acknowledge support through national funds by FCT – Fundação para a Ciência e a Tecnologia, I.P., Portugal, in the framework of the projects the Center for Astrophysics and Gravitation (CENTRA/IST/ULisboa) a through grants No. UID/PRR/00099/2025 and No. UID/00099/2025.

\end{acknowledgements}

\bibliographystyle{aa}
\bibliography{biblio}

\appendix
\section{Relativistic effects and R{\o}mer's delay}
\label{app:rel_effects}

In order to produce a better fit, there are observational effects that must be included in the model. 

The R{\o}mer's delay is the difference between the time of emission of the signal $t_{\rm em}$ and the actual observational dates $t_{\rm obs}$, due to the finite speed of light. To include this delay, we used the first order Taylor's expansion of the R{\o}emer equation, which reads:
\begin{equation}
    t_{\rm em} = t_{\rm obs} - \frac{z_{\rm obs}(t_{\rm obs})}{1 + v_{z_{\rm obs}}(t_{\rm obs})} \,.
    \label{t_em}
\end{equation}
The R{\o}mer effect affects both the astrometry and the spectroscopy, and for S$2$ it has an impact of $\approx 450 \, \mu$as on positions and $\approx 50$ km/s at periastron on radial velocities. 

Moreover, there are two relativistic effects that must be taken into account when the stars approach their periastron: the relativistic Doppler shift and the gravitational redshift. Both induce a shift in the spectral lines that affects the radial velocity measurements. 
The former is given by
\beq 
1 + z_{D} = \frac{1 + v_{z_{\rm obs}}}{\sqrt{1- v^2}} \,,
\eeq 
while the gravitational redshift is defined as 
\beq 
1 + z_{\rm G} = \frac{1}{\sqrt{1 - 2 U(r_{\rm em})}}\,,
\eeq
where $U(r_{\rm em})$ is the potential in Eq.~\eqref{yukawa} evaluated at the time of emission $t_{\rm em}$. 

The two shifts can be combined using Eq.~(D.13) of \citet{Grould:2017bsw} to obtain the total radial velocity 
\beq
V_R \approx \frac{1}{\sqrt{1 - \epsilon}} \cdot \frac{1 + v_{z_{\rm obs}}/\sqrt{1-\epsilon}}{\sqrt{1 - v^2/(1 - \epsilon)}} - 1 \,.
\eeq
where $\epsilon = 2U({r_{\rm em}})$. 

In the total space velocity $v = |\textbf{v}|$ we must also add a correction due to the Solar System motion. We followed the most recent work of \citet{2020ApJ...892...39R} and take a proper motion of \sgra of 
\beq
\begin{split}
& v_x^{\rm SSM} = -5.585 \, \rm mas/yr = 6.415 \cos(209.47^{\circ}) \, mas/yr \, ,\\
& v_y^{\rm SSM} = -3.156 \, \rm mas/yr = 6.415 \sin(209.47^{\circ}) \, mas/yr \, .
\end{split}
\eeq

\section{Coordinate transformation}
\label{app:coord_transf}
The transformation between the co-moving frame of the star $(\mathbf{n}, \mathbf{l}, \mathbf{z})$ and the observer one, identified by $\{\mathbf{X}, \mathbf{Y}, \mathbf{Z}\}$ can be done following \citep{PoissonWill2012}, 
\beq
\begin{bmatrix}
X \\
Y\\
Z\end{bmatrix} = r \begin{bmatrix}
\cos \Oorb \cos(\oorb +f) -\cos \iorb \sin \Oorb \sin (\oorb + f) \\
\sin \Oorb \cos(\oorb +f) +\cos \iorb \cos \Oorb \sin (\oorb + f)\\
\sin \iorb \sin (\oorb + f) \end{bmatrix},
\eeq
where we have used that again $r = p/(1 + e \cos f)$, while
\begin{align}
v_X &= -\sqrt{\frac{G M_{\bullet}}{p}} \left\{ \cos \Oorb \left[\sin (\oorb + f) + e \sin \oorb\right] \right. \notag \\
    &\quad \left. + \cos \iorb \sin \Oorb \left[\cos (\oorb + f) + e \cos \oorb \right] \right\}, \\
v_Y &= -\sqrt{\frac{G M_{\bullet}}{p}} \left\{ \sin \Oorb \left[\sin (\Oorb + f) + e \sin \oorb\right] \right. \notag \\
    &\quad \left. - \cos \iorb \cos \Oorb \left[\cos (\oorb + f) + e \cos \oorb \right] \right\}, \\
v_Z &= \sqrt{\frac{G M_{\bullet}}{p}} \sin \iorb \left[\cos (\oorb + f) + e \cos \oorb\right].
\end{align}

\section{Initial values of the fitting routine}
\label{app:initial_values}

In Table~\ref{initial_values} we report the initial guesses for the shared parameters and the orbital parameters of each star. The initial value of $\alpha$ is generally set to zero, although it was necessary to change it and test different initial values in regimes of strong degeneracy with $M_{\bullet}$ ($\lambda \gtrsim 10^{15}\, \rm m$).

\begin{table}
\caption{Initial parameters used for the MCMC algorithm}
\begin{tabular}{|cc|}
    \hline
    Parameter & Initial value \\
    \hline
    $ M_{\bullet} \, \rm [10^6 \, M_{\odot}]$ & 4.29 \\
    $ R_0 \, \rm [kpc]$ & 8.28 \\
    $ x_0, y_0 \, \rm [mas]$ & 0 \\
    $ v_{x_0}, v_{y_0} \, \rm [mas/yr]$ & 0 \\
    $ v_{z_0} \, \rm [km/s]$ & 0 \\
    \hline
    \multicolumn{2}{|c|}{S2}\\
    \hline
    $ e $ & 0.884 \\
    $ p \, [\rm as]$ & 0.027 \\
    $ i \, [^{\circ}]$ & 134.8 \\
    $ \omega \, [^{\circ}]$ & 66.2 \\
    $ \Omega \, [^{\circ}]$ & 228.2 \\
    \hline
    \multicolumn{2}{|c|}{S29} \\
    \hline
    $ e $ & 0.969 \\
    $ p \, [\rm as]$ & 0.024 \\
    $ i \, [^{\circ}]$ & 144.5 \\
    $ \omega \, [^{\circ}]$ & 205.0 \\
    $ \Omega \, [^{\circ}]$ & 6.2 \\ 
    \hline
    \multicolumn{2}{|c|}{S38} \\
    \hline
    $ e $ & 0.815 \\
    $ p \, [\rm as]$ & 0.047 \\
    $ i \, [^{\circ}]$ & 166.8 \\
    $ \omega \, [^{\circ}]$ & 27.1 \\
    $ \Omega \, [^{\circ}]$ & 109.4 \\ 
    \hline
    \multicolumn{2}{|c|}{S55} \\
    \hline
    $ e $ & 0.737 \\
    $ p \, [\rm as]$ & 0.049 \\
    $ i \, [^{\circ}]$ & 161.2 \\
    $ \omega \, [^{\circ}]$ & 311.3 \\
    $ \Omega \, [^{\circ}]$ & 303.1 \\
    \hline
    \multicolumn{2}{|c|}{S301} \\
    \hline
    $ e $ & 0.984 \\
    $ p \, [\rm as]$ & 0.002 \\
    $ i \, [^{\circ}]$ & 124.0 \\
    $ \omega \, [^{\circ}]$ & 293.4 \\
    $ \Omega \, [^{\circ}]$ & 73.8 \\ \hline
\end{tabular}
\label{initial_values}
\end{table}

\section{Limitations of real-only S301 data for small-$\lambda$ Yukawa constraints}
\label{app:mock_vs_real}

In this Appendix we assess the constraining power of the real S301 data alone (which span the period 2023.25-2025.36) on the Yukawa intensity $\alpha$ at small lengths $\lambda$. The periastron passage of S301 is $t_p \approx 2023.13$ \citep{GRAVITY:2026}, roughly $0.1 \, \rm yr$ before the first observed epoch. So the dataset comprises only observations of the star receding from periastron. 

The exponential term associated with the Yukawa correction is appreciable only when $r$ is smallest, i.e. near the periastron $r_p \approx 10^{13} \, \rm m$. 
For a very highly eccentric orbit like S301's, this confines the information on $\alpha$ in a very narrow window, which current data do not completely cover.

Although the first observed epoch lies quite close to periastron, it is a single data point, which is not sufficient to sample the Yukawa correction where it is largest. In order to constrain it, one requires multiple observations both before and after the pericenter passage, such that the degeneracy with the central BH mass can be broken. 

\begin{figure}
   \centering  \includegraphics[width=\hsize]{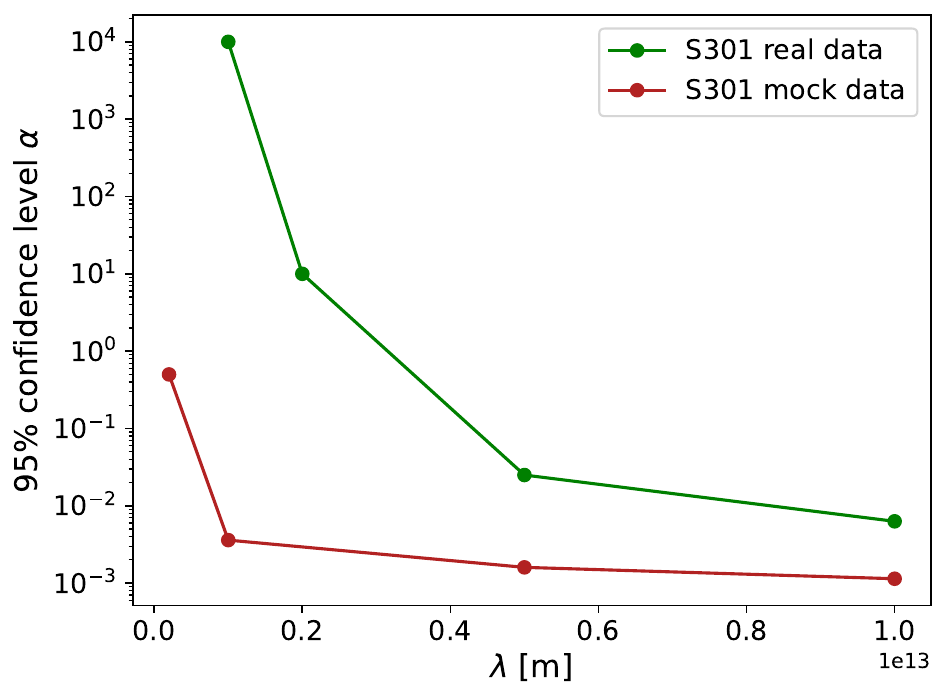}
      \caption{Comparison between $95\%$ confidence level on $|\alpha|$ obtained with mock data (red line, this work) and with only real data for S301 (green line). Other stars are also included. Since we only have post pericenter passage observations, the constraints quickly degradate already at around $\lambda \sim 2 \cdot 10^{12}\, \rm m$, compared with when the pericenter region is covered by mock observations. }
         \label{fig:conf_int_mockvsreal}
\end{figure}

In Figure \ref{fig:conf_int_mockvsreal} we show how the constraints on $|\alpha|$ change if only real data are used, compared to the analysis reported in the main text where the pericenter passage is sampled with mock observations. 

\begin{figure*}
   \centering  \includegraphics[width=\textwidth]{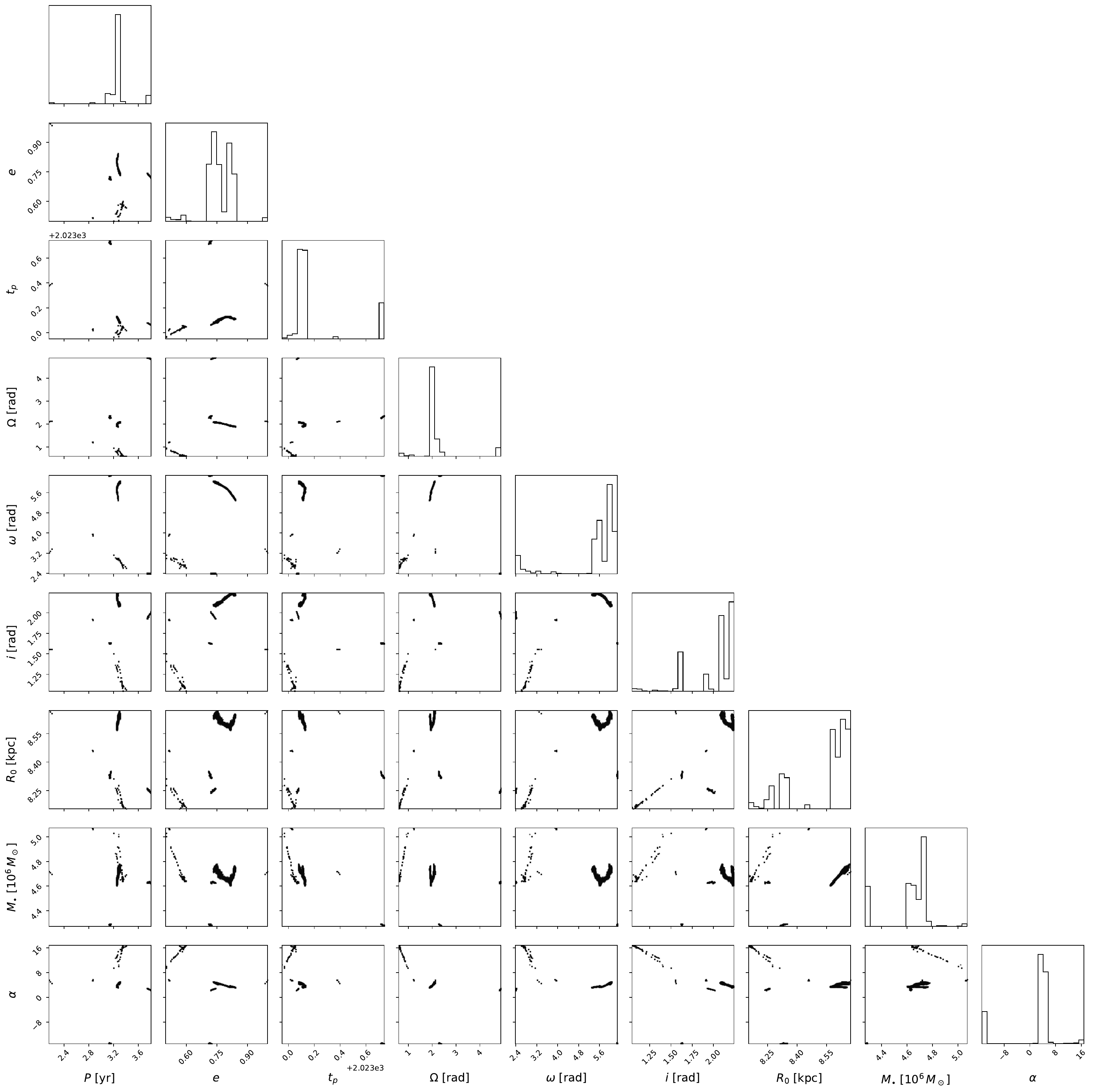}
      \caption{Corner plot showing the correlations between S301 orbital parameters, the common parameters $M_{\bullet}$ and $R_0$ and the Yukawa intensity $\alpha$, when $\lambda = 2 \cdot 10^{12}\, \rm m$ and only real data are used. One can see that, although the fit has formally converged, the posteriors are poorly defined, showing multiple peaks and large uncertainties.}
         \label{fig:corner_plot_uncon}
\end{figure*}

Thus, when using only real data, the constraint on $\alpha$ starts to deteriorate already at $\lambda = 2 \cdot 10^{12} \, \rm m$, where we find a very weak bound of $\alpha \lesssim 10$ at $95\%$ confidence level, roughly three orders of magnitude worse than when mock, pre-pericenter data are included. As explained in the main text, this also implies a degeneracy with the central mass $M_{\bullet}$, which also appears poorly constrained, as well as the other orbital parameters of the star. These features are shown in the corner plot of Figure \ref{fig:corner_plot_uncon}. 

\begin{figure*}
   \centering  \includegraphics[width=\textwidth]{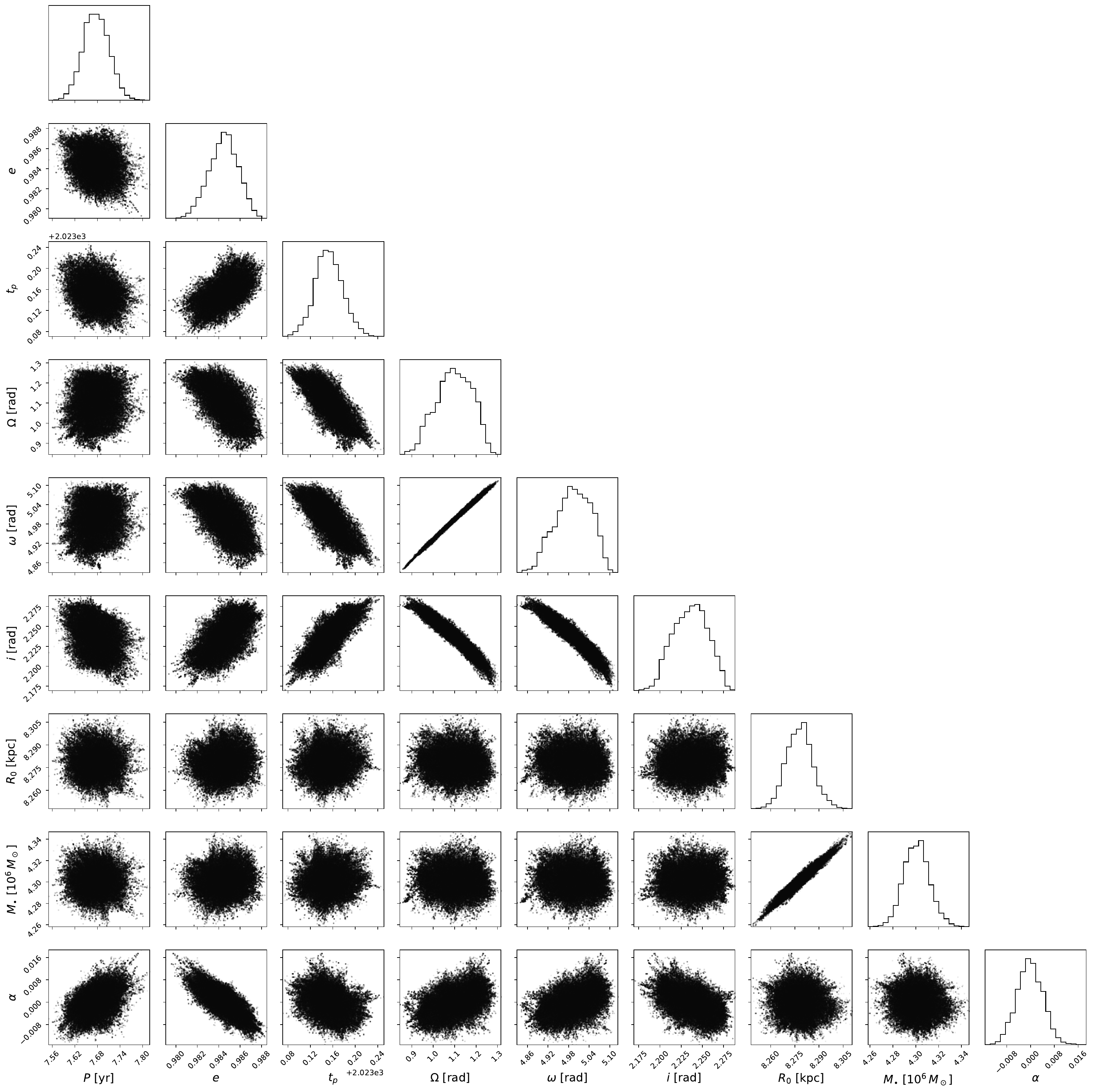}
      \caption{Corner plot showing the correlations between S301 orbital parameters, the common parameters $M_{\bullet}$ and $R_0$ and the Yukawa intensity $\alpha$, when $\lambda = 2 \cdot 10^{12}\, \rm m$ and mock data, and specifically pre-pericenter passage epochs, are used. $\alpha$ is well constrained to be $|\alpha| \lesssim 10^{-2}$. }
         \label{fig:corner_plot_con}
\end{figure*}

For comparison, in Figure \ref{fig:corner_plot_con} we show the corner plot for the same value of $\lambda$ when the mock dataset is used, that is the main analysis reported in this work, which includes pre-periastron epochs. 

\end{document}